\documentstyle{aastex}
\begin{document}

\title{The First Direct Spectroscopic Detection of a White Dwarf Primary
in an AM CVn System}

\author{Edward M. Sion}
\affil{Dept. of Astronomy \& Astrophysics,
Villanova University,
Villanova, PA 19085,
emsion@ast.vill.edu}

\author{Jan-Erik Solheim}
\affil{Institute of Theoretical Astrophysics
University of Oslo,
N-0325 Blindern-Oslo, Norway
jan-erik.solheim@phys.uit.no}

\author{Paula Szkody}
\affil{Department of Astronomy,
University of Washington,
Seattle, WA 98195,
szkody@astro.washington.edu}

\author{Boris T. Gaensicke}
\affil{Department of Physics,
University of Warwick,
Coventry, CV4 7AL, United Kingdom
boris.gaensicke@warwick.ac.uk}

\author{Steve B. Howell}
\affil{NOAO, 950 N. Cherry St.
,Tucson AZ 85726,
Howell@noao.edu}

\begin {abstract}

We report the results of a synthetic spectral analysis of
Hubble STIS spectra of the AM CVn-type cataclysmic variable
CP Eri obtained when the system was in quiescence. The FUV
spectrum is best fitted by a helium-dominated, hybrid
composition (DBAZ) white dwarf with T$_{eff} \sim
17,000K\pm1000$K, log g$\sim8$, He abundance $\sim 1000
\times$ solar, H abundance $\sim 0.1 \times$ solar,
metallicity Z $\sim 0.05 \times$ solar, $V \sin i \sim 400$
km/s$\pm100$ km/s. This is the first directly detected
primary white dwarf in any AM CVn and the surface abundance
and rotation rate for the white dwarf primary are the first
to be reported for AM CVn systems. The model-predicted
distance is $\sim 1000$ pc. The spectral fits using pure He
photospheres or He-rich accretion disks were significantly
less successful. Based upon the analysis of our FUV spectra,
CP Eri appears to contain a hybrid composition DBAZ white
dwarf with a metallicity which sets it apart from the other
two AM Cvn stars which have been observed in quiescence and
are metal-poor. The implications of this analysis for
evolutionary channels leading to AM CVn systems are
discussed.

\end{abstract}


\section{Introduction}

The AM CVn objects, like the essentially pure helium DB
white dwarfs, are very H-poor and appear to be dominated by
nearly pure helium accretion disks in the optical during
outburst and (for those which have low states), in
quiescence. In their bright states they are
spectroscopically similar to the spectra of AM CVn in its
continual bright state in which the absorption lines of an
optically thick helium disk and wind dominate their optical
and FUV spectra (e.g., Groot et al. 2001 and references
therein).

AM CVn objects are widely regarded as interacting binary
white dwarfs in which the less massive degenerate companion
(with M$_{wd} < 0.1 M_{\sun}$) fills its Roche lobe and
transfers He-rich gas through an accretion disk to the more
massive companion white dwarf primary. The basic model was
first proposed by Paczynski (1967), Faulkner, Flannery and
Warner (1972)  with their binary nature first confirmed by
Nather, Robinson and Stover (1981). The properties of these
systems are comprehensively reviewed by Warner (1995).

In addition to the interest in the evolutionary history
which leads to their nearly pure helium composition and the
accretion physics and physical conditions which prevail
during helium accretion, these objects may contribute up to
25\% of the Type Ia supernova production rate (Nelemans et
al. 2001) although recent work suggests it is less than 1\%
(Solheim \& Yungelson, 2005)

The best distance estimates for AM CVns are based on the
Hubble distances given by P.Groot (Nijmegen workshop on AM
CVn stars, July 2005). For the two systems, CR Boo and V803
Cen, most similar to CP Eri, i.e. having outburst/quiescence
states and similar orbital periods, M$_{v}$ = 6.5 for CR Boo
and 5.4 for V803 Cen (for their high states). If we assume
for CP Eri an average M$_{v} = 6\pm 0.5$ for the high state,
we get a distance $1.2\frac{+0.8}{-0.4}$ kpc.

The systems CR Boo, CP Eri, V803 Cen, 2QZJ142701.6-012310,
KL Dra, and SN2003aw are the only systems of the roughly
dozen known AM CVn objects to have both high states and low
states in analogy with their H-rich dwarf novae and
nova-like counterparts. CP Eri, the topic of this Letter,
has an orbital period of 28.73 minutes, an optical
brightness range of 16.5 in outburst and 19.7 in quiescence
with broad shallow optical He I absorption in outburst and,
in quiescence, double-peaked He I and He II emission as well
as Si II emission in the optical (Abbott et al. 1992).
Abbott et al. (1992) did not detect H in either high or low
state optical spectra.

A difficult obstacle to gaining information on the
underlying stars has been that the quiescent spectra are
rare and of poor signal-to-noise due to the objects'
faintness both in the optical and the far UV. In this
Letter, we report the analysis of the first FUV spectrum of
a quiescent AM CVn that displays substantial continuum flux.  
All other AM CVn systems that display substantial continuum
flux are in outburst like AM CVn itself.

\section{Hubble STIS Observation}

We obtained two HST STIS spectra of CP Eri on 11 September,
2000 with the STIS FUV/MAMA configuration and the G140L
grating through the $52\times 0.2$ aperture with exposure
times of 1919 s and 2580 s for spectra O5B606010 and
O5B606020 respectively. The STIS CCD acquisition image
obtained immediately before the G140L spectrum was used to
measure the optical brightness of CP\,Eri during the HST
observations. The image was obtained with the
F28$\times$50LP filter, which has a pivot wavelength at
7229\,\AA, and a band width of 5400--10\,000\,\AA, roughly
comparing to an $R$-band filter (see Araujo-Betancor et al.
2005 for details of the procedure).

Unlike the optical spectrum of CP Eri seen in quiescence,
the FUV spectrum contains many observed absorption features
including a strong feature at Lyman Alpha, strong C III
(1175), C II (1335), Si II (1260, 1265), C I (1270), OI + Si
III (1300), C II (1335), C I (1356, 1490, 1657), Si II
(1526, 1533)  and moderately strong emission features at Si
IV (1393, 1400), possible N V (1238, 1242) emission, and
possible He II (1640) emission.

\section{Synthetic Spectral Fitting with Helium Disks and Photospheres}

The grid of helium accretion disk models described by Nasser
et al.  (2001) were compared with the HST STIS spectrum of
CP Eri over the effective wavelength range of the STIS
spectrum, 1150\AA -- 1716\AA.  The helium accretion disks
are steady state NLTE models which are more appropriate for
the high state of AM CVn systems. However, as a first
approximation to the accretion disk during a low state, we
applied the optically thick models to CP Eri's spectrum.

The composition and designation of the disk models (see
Table 1 below) is as follows. The disk models labeled
cperim4i* have He/H =1000, Z = 0.001 solar, and an outer
radius (of the outermost annulus) of 15 white dwarf radii.
The "{\it i}" is the inclination angle in degrees. The disk
model cperim9i45 has He/H = 105, CNO abundances = 3, 900,
$1.5\times$ solar respectively, and outermost disk radius =
15 WD radii. The disk model cperim10i* has He/H = 105,
metallicity Z = solar and outer disk radius r$_{max} = 8$ WD
radii. Curiously, the disk fits are improved considerably
below 1350AA if Fe is overabundant because of the large
number of low exitation Fe lines whose collective absorption
eats away at the continuum and broaden line profiles. This
is especially noticeable at 1260AA. On the hand, when the
same disk models are applied to AM CVn itself, the elevation
of Fe does not result in a better fit.  It is possible, this
difference between CP Eri and AM Cvn may point to different
progenitor evolution.

In preparation for the model fitting, emission lines in the
data were masked out. We used a $\chi^{2}$ minimization
fitting routine wddiskfit which yields the $\chi^{2}$ value,
scale factor and the distance computed from the scale
factor.  We have tabulated the results in Table 1 where the
first column lists the model designation (see above), second
column the $\chi^{2}$ value third column the scale factor,
the last column the distance in kiloparsecs.  To obtain the
distances from the model normalization for each fit, we
scaled down by a factor corresponding to the magnitude
difference between the high state (16.5) and the low state
(19.7). This corresponds to a factor of 0.0524 or a distance
ratio 0.229. This yields distances of 1.34, 1.04, 1.21,
1.27, 1.15 kpc, which are all quite reasonable. The
best-fitting accretion disk model is cperi4i45, which has
He/H = 1000, Z=0.001(including Fe), a disk inclination angle
of 45 degrees and a $\chi^{2} = 2.4162$. The "best-fit"  
helium accretion disk fit is displayed in figure 1.
                                                        
\begin{deluxetable}{lccc}  
 \tablecaption{Helium Accretion Disk Model Fits}
\tablehead{
\colhead{                                                         
  Model}&\colhead{$\chi^{2}$}&\colhead{Scale Factor}&\colhead{Distance(kpc)}}
\startdata
cperim4i30 &                    3.15  &      $2.91\times 10^{-04}$&        1.34 \\     
cperim4i45  &                   2.42  &$  4.82\times 10^{-04}$&        1.04\\
cperim9i45  &                   3.12    &$    3.56\times 10^{-04}$&        1.21\\
cperim10i30  &                  4.52   &$     3.26\times 10^{-04}$&        1.27\\
cperim10i45   &                 4.54     &$   3.99\times 10^{-04}$&        1.15\\
\hline
\enddata
\end{deluxetable}

We also explored the possibility that the STIS spectrum of
CP Eri in its low brightness state, like the FUV spectra of
the shortest period dwarf novae, is produced by a white
dwarf with essentially no contribution from an accretion
disk. Therefore, we constructed an initial grid of
helium-rich photospheres with He/H $= 10^{-5}$, He = 1000,
and Z $= 10^{-4}$.  The grid covers the following parameter
ranges: temperatures of 15,000K - 30,000K in steps of 3000K,
surface gravities log g = 7.5, 8.0, 8.5 and rotational
velocities $V \sin i = 200$, 400 and 600 km/s. The
best-fitting helium-rich photosphere model has T$_{eff} =
15,000$K and log g = 8.0. The rotational velocity is
meaningless since the low metallicity model had no strong
metal absorption lines to match with the STIS spectrum.  
This model yielded a $\chi^{2} = 2.5702$, a scale factor $=
3.29 \times 10^{-4}$ and a distance of 804 pc for a white
dwarf radius R$_{wd}$/R$_{\sun} = 1.46 \times 10^{-2}$.  
The best-fitting He photosphere (no H) is displayed in
figure 2. However, this model does not fit the absorption
lines well.

The rather deep absorption line near 1216\AA\ could not be
due to He II at the T$_{eff}$ of the white dwarf indicated
by the continuum and by low ionization metal line profile
fits. Hence, unless the absorption has a hydrogen
interstellar origin, which is unlikely given its breadth,
there is a possibility it is photospheric H I Lyman$\alpha$.
Therefore, we explored hybrid composition "DBA" atmospheres
in which the dominant element is helium with hydrogen being
far less abundant. Assuming that the profile is entirely H
I, we kept the gravity fixed at log g = 8 and experimented
with various He/H ratios from 102 to 105, metal abundances Z
$= 0.5, 0.1, 0.05, 0.005$, T$_{eff} = 14,000,
15,000...,20,000$K. We found that the optimal He/H ratio
needed to replicate the profile is He = 1000, H = 0.1 or
He/H = 10000.  This ratio is smaller than the stringent He/H
ratio characterizing the DB white dwarfs where He/H $> 105$
in order for Balmer lines not to be detected in their
optical spectra (which they are not). The best-fitting
hybrid composition "DBA"  model had the following
parameters: $\chi^{2} = 1.45$, scale Factor S $= 2.224
\times 10^{-4}$, Log g = 8 (fixed), T$_{eff} = 17,000$K, $V
\sin i = 400$ km/s, He $= 10^{3}$, H = 0.1 and Z = 0.05 and
a model-predicted distance of 978 pc. This best-fit model,
compared with the STIS data, is shown in figure 3. The
hybrid atmosphere fit (H + He) provides a reasonably good
fit to both the continuum and the absorption line profiles.
The Lyman$\alpha$ absorption profile, assuming no part of it
is interstellar, is fit very well with the chosen mix of H
and He. However, at the metal abundance of 0.005 solar,
while the Si II features at 1260, 1265 are quite well-fit
along with C II (1335) and Si II (1526, 1533), the C III
(1175), S III + OI (1300) and C I (1356, 1657)  absorption
features are not well-fit by the model with the synthetic
profiles, being considerably weaker than the observed ones.
Still, we are encouraged that at least for the Lyman$\alpha$
profile, and the lower ionization lines of C and Si, the fit
appears to be somewhat consistent. This model, compared with
the best-fit disk model, has a lower $\chi^{2}$, and fits
the metal lines and the Lyman alpha region successfully
whereas the best-fit disk model fails to do this.

An additional test of the consistency of our DBAZ
composition WD fit is offered by the constraint that the
magnitude corresponding to the optical or IR flux of the
model is not brighter than the corresponding observed
magnitude of CP Eri in the same wavelength range, since it
is expected that other sources of systemic light (e.g. a hot
spot, accretion disk, secondary) are contributing to the
system brightness as well. The STIS F28x50LP magnitude at
the time of the HST observation was 19.9. Our best-fit white
dwarf model folded with the acquisition filter transmission
predicts a magnitude of 20.8, fainter than the observed
value.

\section{Discussion and Conclusions}

Our analysis suggests that CP Eri may not be a typical AM
CVn system in that we find a significant abundance of H and
a higher metallicity compared with other AM CVn systems such
as the well-studied object, GP Com which is metal poor, has
shown little evidence of any H and is always seen in a low
state. It is obviously important to explore whether the
abundance of H from our UV analysis would lead to detected H
features in the optical spectrum. While a re-examination of
the optical quiescent spectrum of Groot et al. (2001),
suggests a possible hint of very weak H I emission features
in the optical low state spectrum (see figure 1 in Groot et
al. 2001), much higher signal to noise optical spectra are
clearly needed. In any case, the metallicity we derive is
consistent with the Groot et al. conclusion that CP Eri has
higher metallicity than GP Com and CE315, implying that it
is not a population II object.

Is it reasonable for the white dwarf in an AM CVn star like
CP Eri to be accreting both He and H? If so, what are the
implications for the ancestry of CP Eri and other AM CVn
objects? There are currently three formation channels
favored for AM CVn stars: (1) the double degenerate scenario
(Tutukov and Yungelson 1979); (2) semi-degenerate helium
star scenario (Iben and Tutukov 1991); and (3) subset of
H-rich CVs with evolved secondaries (Podsiadlowski et
al.2003). In the latter scenario, a normal H-rich star of
mass $\sim 1 M_{\sun}$ fills its Roche lobe near the end of,
or just after, core Hydrogen burning while the initially
non-degenerate and H-rich companion becomes increasingly
helium-rich and degenerate during its evolution. In their
early evolution, these systems would appear as "normal"
H-rich CVs with evolved secondaries. A number of theoretical
investigations of all three formation channels have been
carried out with stellar evolution codes and binary
population synthesis simulations (e.g. Nelemans, G.,
Portegies, Zwart, S.F., Verbunt, F., \& Yungelson, L.R.2001;
Podsiadlowski, Han, and Rappaport, S. 2003). Binary stellar
evolution model sequences using a Henyey-type code and
including angular momentum losses due to magnetic braking
and gravitational wave emission are available for different
evolutionary phases of the evolved donor at the onset of
mass transfer (Podsiadlowski et al. 2003). Only two of their
four evolutionary sequences reach orbital period minima
shorter than 55 minutes (75 minutes is the P$_{orb}$ minimum
for an H-rich CV). The two sequences correspond to donors
with H-exhausted cores of 0.037 and 0.063 M$_{\sun}$ which
both transform themselves into nearly pure He degenerates
but with a few per cent traces of H remaining.  Both of
these sequences pass through the range of AM CVn periods
twice, once with the period decreasing toward the minimum,
once after the period minimum.

Which case might be applicable to CP Eri? Since the
degenerate donors in systems before the period minimum have
larger amounts of H left in their envelopes, it seems more
likely this case would better apply to CP Eri. For its
observed P$_{orb}$ (28.73 minutes), the binary population
synthesis calculations of Podsiadlowski et al.(2003) yield
predicted values of secondary mass M$_{2} = 0.100$
M$_{\sun}\frac{+0.003}{-0.021}$, mass transfer rate \.{M}$ =
10^{-9.4}$ M$_{\sun}$/yr and surface hydrogen abundance on
the secondary, X = 0.22 for CP Eri if its P$_{orb}$ is
decreasing (i.e., it is evolving before the period minimum).
If however, CP Eri is evolving after the period minimum
(i.e., P$_{orb}$ is increasing), then the simulations
predict M$_{2} = 0.040$ M$_{\sun}\frac{ +0.005}{-0.004}$,
\.{M}$ = 10^{-9.8}$M$_{\sun}$/yr, and X = 0.03
M$_{\sun}\frac{+0.01}{-0.03}$.

Since CP Eri's accretor is the only white dwarf in an AM CVn
so far with a directly determined photospheric H abundance,
we are unable to draw any comparisons with other AM CVn
cases. Therefore, analyses of other exposed white dwarfs in
these objects are clearly needed.

\acknowledgements

We would like to thank Ivan Hubeny for the program Tlusdisk187
and for his help and encouragement in using it.
It is a pleasure to thank Lev Yungelson for a useful
discussion of AM CVn formation channels. One of us (EMS)
would like to acknowledge the kind hospitality of the
Institute of Theoretical Astrophysics at the University of
Oslo where part of this work was carried out. This work was
supported by NASA HST grant GO-8103.01 and in part by NSF
grant AST05-07514. BTG was supported by a PPARC Advanced
Fellowship.

Figure Captions

Fig. 1 - The flux distribution, flux versus wavelength, for
the best-fitting helium accretion disk model with Z = x.x,
and inclination i = 45 degrees compared with the HST STIS
spectrum of CP Eri. \medskip

Fig. 2 - The flux distribution, flux versus wavelength, for
the best-fitting pure helium photosphere model with log g =
8, T$_{eff} = 15,000$K, Z = 0.05, and $V \sin i = 200$ km/s,
compared with the HST STIS spectrum of CP Eri. \medskip

Fig. 3 - The flux distribution, flux versus wavelength, for
the best-fitting hybrid composition,"DBAZ", photosphere
model with log g = 8, T$_{eff} = 17,000$K, He = 1000, H =
0.1, Z = 0.05, and $V \sin i = 400$ km/s, compared with the
HST STIS spectrum of CP Eri.


\begin{thebibliography}{}
\bibitem[]{}
Abbott, T.M.C., Robinson, E.L., Hill, G.J., \& Haswell, C.,
1992, ApJ, 399, 680

\bibitem[]{}
Araujo-Betancor et al. 2005, ApJ 622, 589
\bibitem[]{}
Faulkner, J., Flannery, B., \& Warner, B. 1972, ApJ, 175, L79

\bibitem[]{}
Groot, P.J., Nelemens, G., Steeghs, D.,\& Marsh, T. 2001, ApJ, 558, L123

\bibitem[]{}
Iben, I., \& Tutukov, A.V. 1991, ApJ, 370, 615

\bibitem[]{}
Nasser, M.R., Solheim, J.-E., \& Semionoff, D. 2001, A\&A, 373, 222

\bibitem[]{}
Nather, E. Robinson, E.L. \& Stover, R. 198 {\bf ?}, ApJ, 244, 269

\bibitem[]{}
Nelemans, G., Portgies Zwart, S.F., Verbunt, F., \& Yungelson, L.R. 2001,
   A\&A, 368, 939

\bibitem[]{}
Paczynski, B. 1967, Acta Astr., 17, 287

\bibitem[]{}
Podsiadlowski, Ph., Han, Z., \& Rappaport, S. 2003, MNRAS, 340, 1214

\bibitem[]{}
Solheim, J.-E., \& Yungelson, L. 2005, ASPS, 334, 387

\bibitem[]{}
Tutukov, A.V., \& Yungelson, L. 1979, Acta.Astr., 29, 665

\bibitem[]{}
Warner, B. 1995, in "Cataclysmic Variables" (Cambridge: Cambridge 
University Press){\bf CORRECT REF?}
\end{thebibliography}
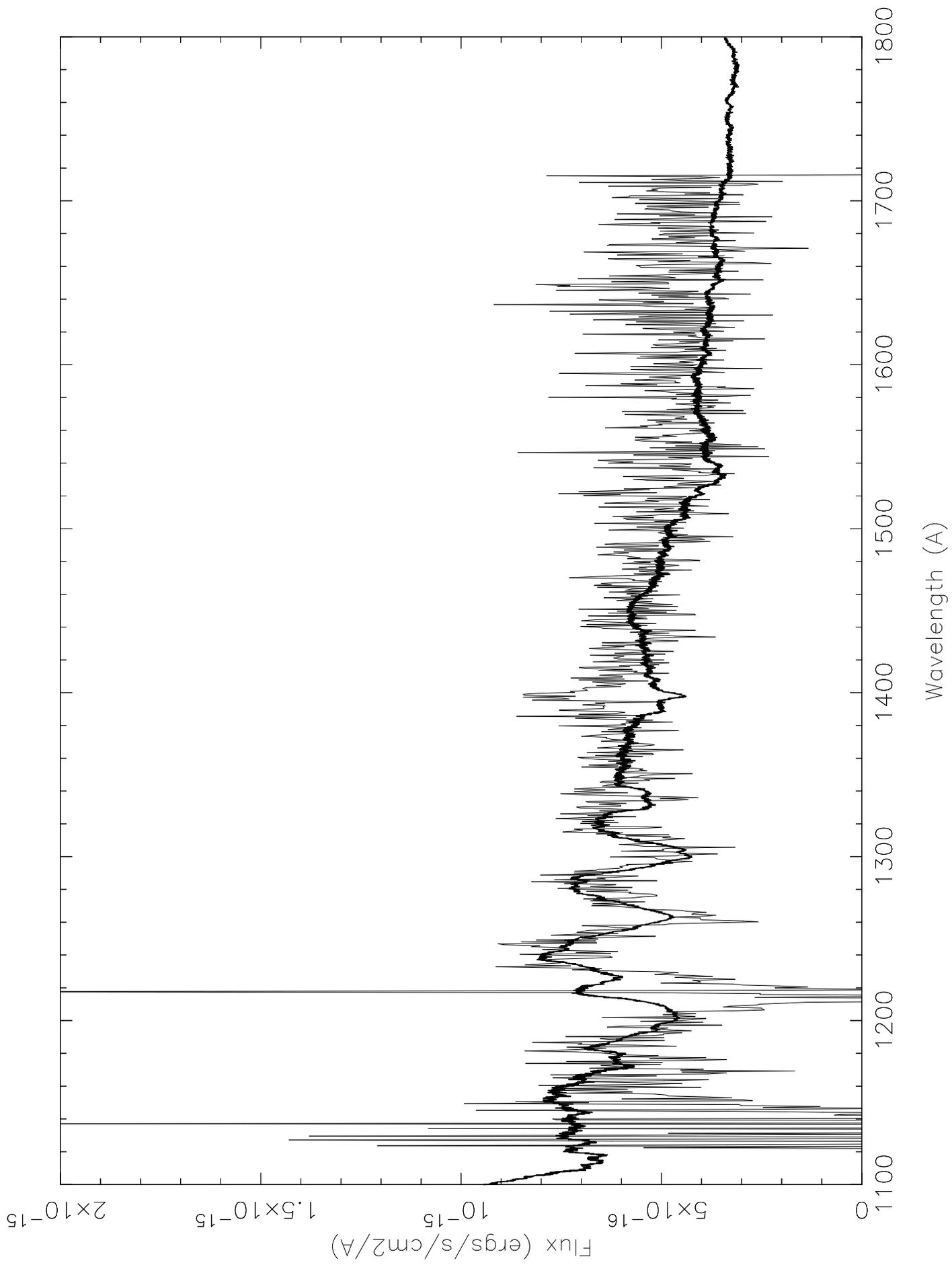
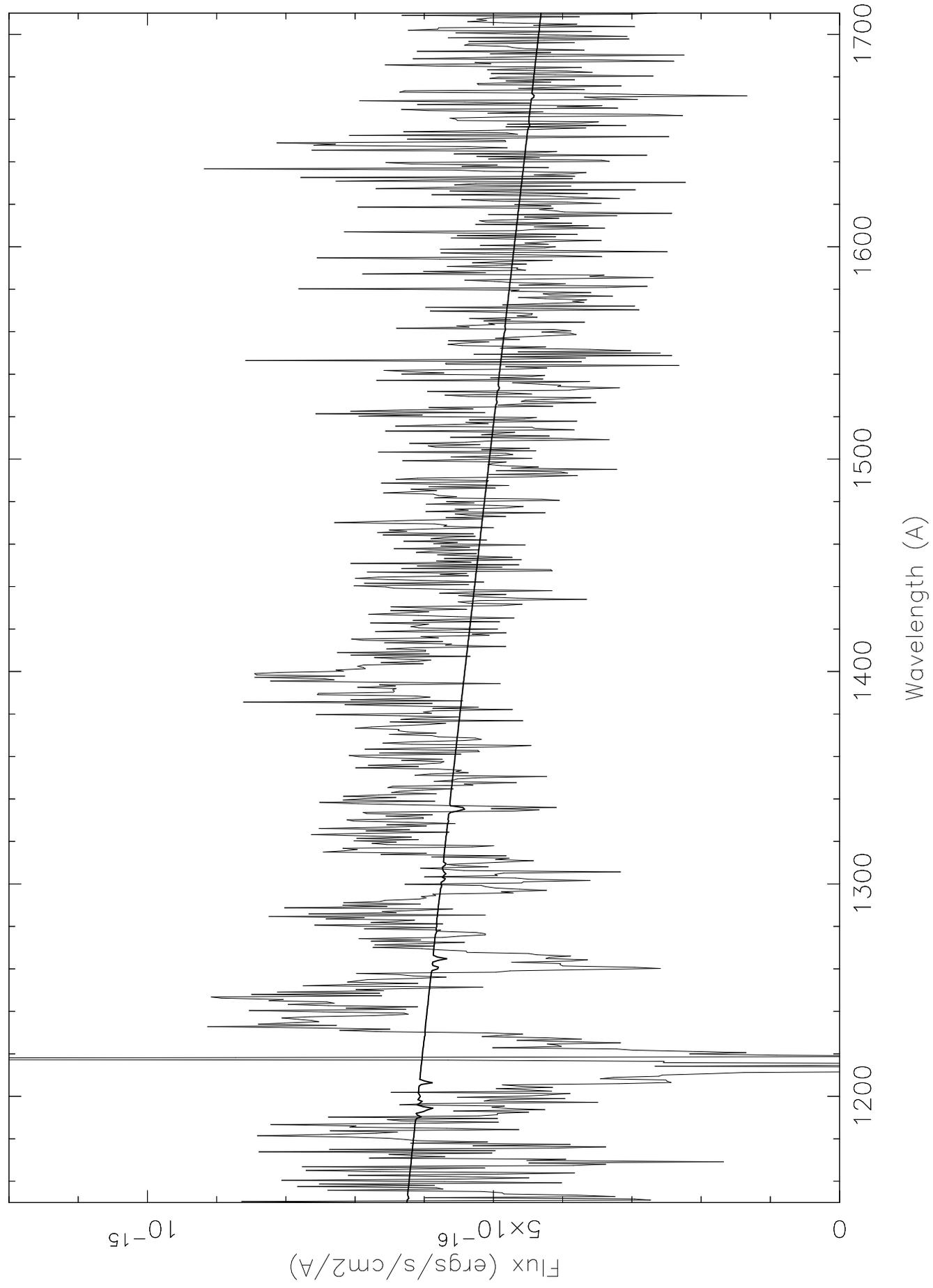
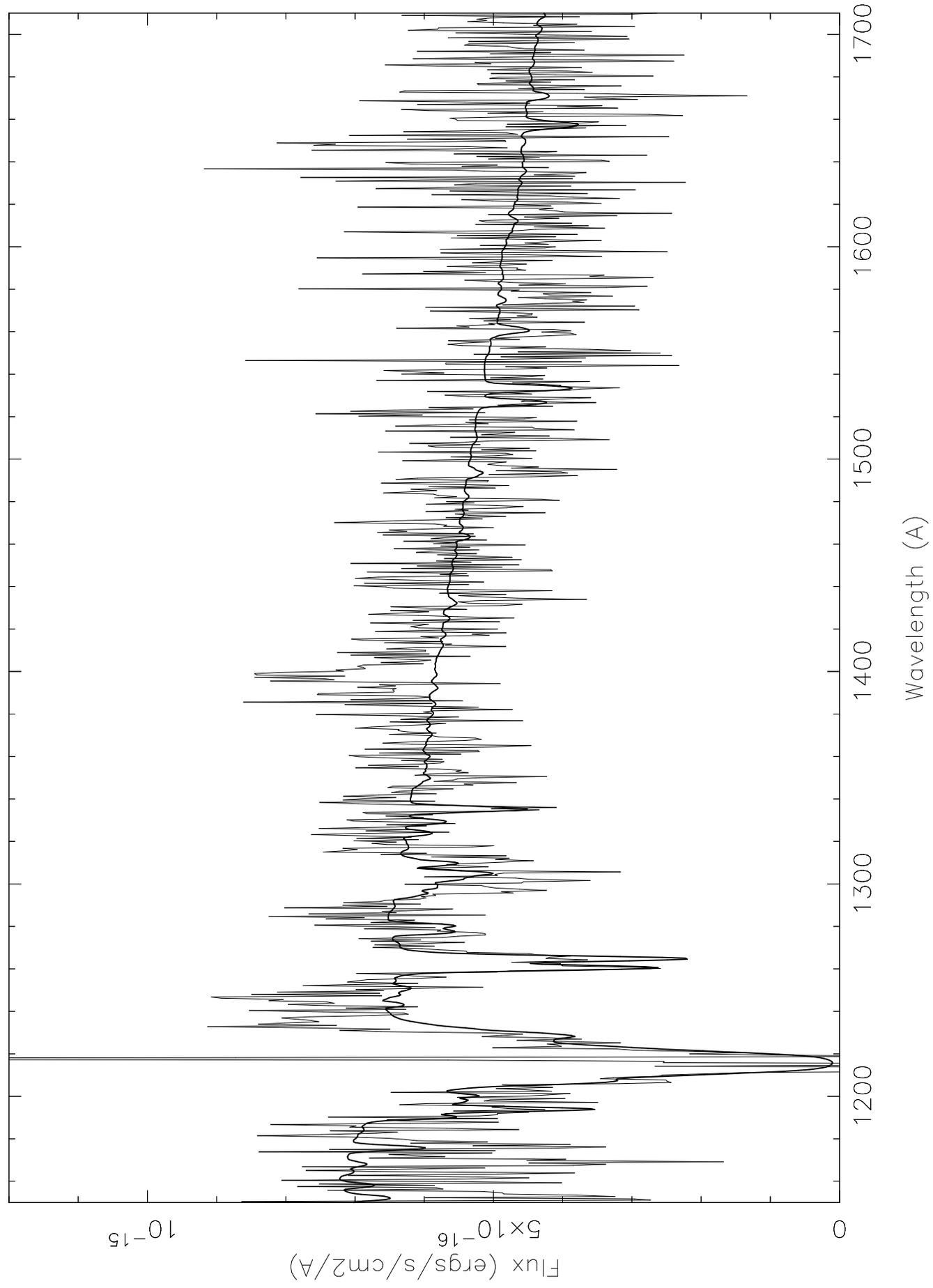
\end{document}